# 人工智能在电力系统安全稳定分析中的应用：综述

张润颢

（湖北工业大学电气与电子工程学院，湖北省 武汉市 430072）

## Artificial Intelligence in Power System Security and Stability Analysis: A Comprehensive Review

ZHANG Runhao

(School of Electrical and Electronic Engineering, Hubei University of Technology, Wuhan 430072, Hubei Province, China)

**ABSTRACT:** This review comprehensively examines the integration of artificial intelligence (AI) in enhancing the dynamic security assessments of modern power systems. It highlights the pivotal role of AI in facilitating scenario generation, incident prediction, risk assessment, and severity grading, thereby addressing the complexities introduced by renewable energy integration and advancements in digital grid technologies. The paper delves into data-driven techniques, with a particular focus on decision trees that effectively bridge operational characteristics with security metrics. These methodologies enable real-time, accurate predictions of system behaviors under varied operational conditions and support the optimization of control strategies. Through detailed analysis, we demonstrate how AI applications can transform traditional security assessment protocols, enhancing both the efficacy and efficiency of power system operations. The findings advocate for the potential of AI to significantly enhance the reliability and resilience of electrical grids, marking a paradigm shift towards more adaptive and intelligent power infrastructure.

**摘要**：本文综述了人工智能（AI）在现代电力系统动态安全评估中的应用，重点探讨了 AI 在运行场景生成、事故预测、风险评估和严重程度分级等方面的关键作用。文章强调，通过利用数据驱动的技术，尤其是决策树方法，AI 技术能够有效地解决由可再生能源集成和数字化电网技术带来的复杂性问题。这些方法不仅支持系统在不同运行条件下的实时、准确预测，还优化了控制策略，从而提升了电力系统操作的效率和效果。本研究通过详尽的分析，展示了 AI 在传统安全评估协议中的转型潜力，增强了电力系统的可靠性和弹性。研究结果支持 AI 技术在电网可靠性和弹性提升中的重要潜力，预示着向更适应性强、智能化的电力基础设施的转变。

**关键词**：人工智能；电力系统稳定性；动态安全评估；决策树；风险管理

## 0 引言

电力系统静态安全分析及预警针对特定预想事故集对未来可能发生的运行状态进行安全性性分析和校验，辨识出不安全运行场景；进而对其不安全程度进行量化评估并划分风险等级；最后通过安全风险分析及预警结果指导后续的安全风险防范控制，是保障电力系统安全稳定运行的关键分析技术之一。由于新能源发电的强不确定性和复杂时空相关性电力系统实际运行中可能出现的运行场景数目呈海量增长[1]。另外，由于电力电子设备控制呈现多时间尺度级联和控制切换的特点[2]，电力系统动态模型的复杂度显著增加。实现快速准确的新型电力系统动态安全风险预警面临更大困难。

由于新能源发电的不确定性和新型电力系统动态模型的高维、强非线性，基于物理模型和时域仿真的安全风险预警方法难以满足在线计算的速度要求。随着以深度学习为代表的新一代人工智能技术的快速发展，数据驱动的安全风险预警方法得到了国内外学者的愈加关注[3-4]。数据驱动的安全风险预警方法利用机器学习模型建立电网

运行特征与功角、电压和频率等不同动态安全风险评估指标之间的非线性映射关系[5-6]。这类方法离线阶段使用大量样本训练机器学习模型，在线阶段利用训练好的模型评估安全风险。数据驱动的安全风险预警方法直接从训练样本中挖掘运行特征与安全风险指标之间的关联关系，不需要求解表征电力系统动态行为的高维非线性微分-代数方程组，能够在保证一定精度的前提下实现动态安全风险的快速评估。

# 1 电力系统静态安全分析框架

电力系统安全风险预警主要包括运行场景生成与缩减、预想事故筛选、安全风险评估和严重度分级4个环节。

运行场景生成与缩减提供未来代表性运行场景及其概率信息，预想事故筛选提供高风险的常规预想事故和连锁故障集。基于典型运行场景和预想事故集，新型电力系统的安全风险评估包括对暂态电压稳定性、同步稳定性和暂态频率安全的定量评估，以输出不同安全风险属性的评估结果。安全风险严重程度按照特定的分级策略将运行场景的不安全性分级，形成不同的预警级别，并将安全风险分级预警结果用于指导后续的预防控制决策和紧急控制策略的制定。

其中运行场景生成通常基于滚动更新的新能源发电和负荷预测信息，并结合发电调度计划，可以生成大量符合新能源出力时空相关性的未来运行场景。为了简化安全风险预警计算，可以从这些运行场景中筛选出少量代表性的场景进行详细计算，并利用它们的计算结果来代表整体的安全风险预警结果，从而减少需要评估的运行场景数量。预想事故的筛选包括从原始预想事故集中挑选出少量高风险事故进行详细分析，涵盖了N-1、N-k等传统预想事故的选择[7]，同时也考虑了连锁故障的筛选。安全风险评估使用时域仿真方法或数据驱动方法，对每个运行场景在特定预想事故下不同安全风险属性进行量化评估。在安全风险评估基础上，通过预设的严重性分级策略确定每个运行场景的严重性等级。

# 2 基于智能算法的预想事故筛选

预想事故的筛选考虑了事故发生的概率和严重后果，将所有预想事故进行排序，并选取高风险预想事故进行详细分析。在此过程中，如果利用数据驱动技术对系统的暂态行为进行快速评估，将有效提升预想事故筛选的速度和效率。在常规预想事故筛选方面，文献[8]通过对电网历史运行场景进行聚类，为每一类场景分别训练支持向量机作为暂态稳定评估模型，以加速预想事故筛选的过程；文献[9]首先构建节点的暂态电压安全性指标，并利用这些指标的余弦距离度量预想事故之间的相似性，通过层次聚类算法进行预想事故的分组，进而从每一组中筛选出严重的预想事故。但该方法中使用时域仿真计算暂态电压安全性指标向量，将影响预想事故筛选的速度。

在连锁故障筛选方面，文献[10]利用堆叠降噪自动编码器评估连锁故障发生后引起直流换相失败的持续时间，从而快速评估连锁故障引发直流连续换相失败的概率；文献[11]使用堆叠降噪自动编码器建立交流故障发生后是否引起直流闭锁的快速判断模型，辅助搜索能够引发直流闭锁的高风险连锁故障。文献[10]和[11]中建立的深度学习评估模型主要依赖系统的网络结构和故障位置信息作为输入特征，但对系统在发电负荷功率注入变化后的影响考虑不足。为了在连锁故障筛选中考虑新能源接入可能带来的源荷不确定性，文献[12]提出了一种结合分布因子的随机响应面法。该方法用于计算线路连锁开断过程中的随机潮流，并利用深度森林技术快速评估随机潮流计算结果的误差水平。

# 3 基于智能算法的静态稳定评估

安全风险评估的核心任务是评估不同安全风险属性在预想事故发生后的量化表现。由于新型电力系统可能会产生多种运行场景，其动态模型具有高维度和强非线性特性，完全依赖时域仿真计算安全风险评估指标将消耗大量时间。数据驱动的安全风险评估利用机器学习技术提高了评估速度。近年来，深度学习、集成学习、增量学习、迁移学习等新一代机器学习技术在新型电力系统的安全风险评估中得到了广泛应用。在此介绍基于决策树的静态电压稳定评估方法。

## 3.1 基于参与因子的特征变量初筛

在电力系统实际运行中，导致电压失稳的主要原因是系统无法满足负荷的无功需求。发电机和负荷分别产生无功和消耗无功，此外，输电系统中的无功损耗对母线电压有显著影响。因此，本文选择母线、发电机以及支路作为影响静态电

压稳定问题的关键组成部分。通过对系统的静态潮流方程线性化可知：

$$\begin{bmatrix} \Delta \boldsymbol{P} \\ \Delta \boldsymbol{Q} \end{bmatrix} = \begin{bmatrix} \boldsymbol{J}_{P\theta} & \boldsymbol{J}_{PV} \\ \boldsymbol{J}_{Q\theta} & \boldsymbol{J}_{QV} \end{bmatrix} \begin{bmatrix} \Delta \boldsymbol{\theta} \\ \Delta \boldsymbol{V} \end{bmatrix} \quad (1)$$

式中，$\begin{bmatrix} J_{P\theta} & J_{PV} \\ J_{Q\theta} & J_{QV} \end{bmatrix}$ 为系统的潮流雅可比矩阵。

由于电压主要受无功功率的影响，假设 $\Delta P = 0$ 可得：

$$\Delta \boldsymbol{Q} = \left[ \boldsymbol{J}_{QV} - \boldsymbol{J}_{Q\theta} \boldsymbol{J}_{P\theta}^{-1} \boldsymbol{J}_{PV} \right] \Delta V \quad (2)$$

简化有：

$$\Delta V = J_R^{-1} \Delta Q \quad (3)$$

式中，$J_R = [J_{OV} - J_{O\theta} J_{P\theta}^{-1} J_{PV}]$ 为降阶雅可比矩阵。

对 $J_R$ 进行分解可得：

$$J_R = \xi \Lambda \eta \quad (4)$$

式中，$\Lambda$ 为特征值 $\lambda_i$ 的对角矩阵；$\eta$ 和 $\zeta$ 分别为 $J_R$ 的左右特征向量矩阵。

将式(4)代入式(3)可得

$$\eta \Delta V = \Lambda^{-1} \eta \Delta Q \quad (5)$$

令 $\Delta V_m = \boldsymbol{\eta} \Delta V, \Delta Q_m = \boldsymbol{\eta} \Delta Q$ 可得

$$\Delta V_{mi} = \frac{1}{\lambda_i} \Delta Q_{mi} \quad (6)$$

式中，$\lambda_i$ 为 $J_R$ 的第 $i$ 个特征值；$\Delta V_{mi}$ 和 $\Delta Q_{mi}$ 分别为 $\lambda_i$ 对应的模态电压及模态无功变化量。

### 3.2 基于决策树的静态电压稳定裕度评估

通过决策树建立用于静态电压稳定评估的分类学习模型，利用系统中大量存储的历史数据驱动模型训练。通过构建的决策树学习模型，可以提取相应的电压稳定状态评估规则，并对系统当前的电压稳定状态进行判断。

使用决策树进行静态电压稳定评估是将其视作一个分类问题。在机器学习的有监督学习领域中，分类问题是一个重要的分支，涵盖了多种技术，如决策树、神经网络、支持向量机和朴素贝叶斯等。相对于其他分类技术，决策树具有白箱模型的特点。它不仅能为运行人员提供可靠的系统电压稳定状态评估结果，还能通过其树形结构展示其学习的整个过程[13]。这种特性不仅提高了分类结果的可信度，同时也为运行人员在控制系统电压稳定状态时提供了有价值的参考和辅助。图 1 为决策树基本结构。

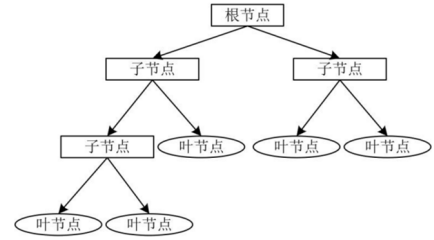

**图 1 决策树基本结构**

如图 2 所示，决策树是一种类似于树的结构，由树节点和树枝组成。树节点包括根节点、子节点和叶节点三种类型。根节点作为起始点存储初始样本集，这些样本具有不同的类别标签，整体上呈现出一定的混乱度。决策树利用熵来衡量样本集的混乱程度，熵值越大，样本集越复杂。

为了降低样本集的熵值，决策树从根节点开始选择一个属性，将初始样本集划分为两个熵值减小的子样本集，并将它们存储在相应的子节点中。子节点接收到样本子集后，继续选择属性进行进一步划分，重复这一过程，将样本子集传递到下一级子节点。

随着样本集不断分裂，每个样本子集中的样本最终具有相同的类别标签，这时熵值为零，将该子样本集存储在叶节点中。从根节点到叶节点的路径完整展示了如何通过选择属性逐步将初始混乱的样本集划分为纯净的子样本集的过程。然后利用 C4.5 算法构建面向电压稳定裕度评估的决策树模型，包括以下四个步骤。

1）初始信息熵。

$$Entropy(S) = -\sum_{i=1}^{m} p_i \log_2 p_i \quad (7)$$

式中：S 为生成样本集；m 表示样本集中样本的类别数目，电压稳定裕度评估中通常被分为正常状态、预警状态和紧急状态 3 类；$p_i$ 表示样本集 S 中属于类别 $i$ 的样本占样本总数的比例。

计算完样本集的初始信息熵后，就需要计算

各个属性对于样本集熵降低的能力，即计算分裂信息熵。

2）属性信息熵。

$$Entropy_A(S) = \frac{|S_L|}{|S|}Entropy(S_L) + \frac{|S_R|}{|S|}Entropy(S_R) \quad (8)$$

式中：A 为任意一个关键特征变量；$S_L$ 和 $R_S$ 是依据属性A将样本集划分后的两个样本子集；$|S_L|$、$|S_R|$ 及 $|S|$ 表示对应样本集中样本的个数。

由上式可以看出，属性信息熵本质上是各样本子集初始信息熵的加权平均值。将样本集的初始信息熵和属性信息熵相减即可量化该属性降低训练样本混乱程度的能力。

3）信息增益。

$$Gain(A) = Entropy(S) - Entropy_A(S) \quad (9)$$

信息增益的大小反映了样本集熵值减少的程度，即该属性在分类中的有效性。ID3 算法利用信息增益来选择节点中最佳的属性，但为了获取最大的信息增益，ID3 算法可能会倾向于将每个样本划分为单独的子样本集。为了克服 ID3 算法可能出现的过拟合问题，C4.5 算法在计算信息增益的基础上引入了分裂信息的概念，从而计算各个属性的信息增益率。这一改进有效地解决了 ID3 算法中可能出现的问题，使得决策树构建过程更加合理和实用。

4）信息增益率。

$$SplitInfor(A) = -\sum_{i=1}^{k}\frac{|S_i|}{|S|}\log_2\frac{|S_i|}{|S|} \quad (10)$$

式中：SplitInfor(A)表示属性 A 的分裂信息。利用属性 A 的信息增益和分裂信息，即可得到相应的信息增益率：

$$GainRatio(A) = \frac{Gain(A)}{SplitInfor(A)} \quad (11)$$

从根节点到最后的子节点，每个节点都会计算所有属性的信息增益率。然后，选择信息增益率最高的属性来对样本集进行划分，这一过程逐级进行，直到样本集被划分为仅包含同一类别样本的纯净子集为止。一旦构建完成决策树，就可以根据决策树的路径提取相应的电压稳定评估规则。系统运行人员可以将 PMU 测量的数据与提取的评估规则进行比对，以评估系统当前的电压状态。

在建立决策树时，只能使用系统中存储的历史电压幅值、历史无功功率和历史断面潮流等过往运行数据。因此，所建立的决策树能够很好地学习和预测这些历史数据。然而，随着电网运行方式的不断复杂化，可能会出现之前未曾遇到的新的运行模式。决策树可能无法通过学习历史数据来有效地应对新的运行模式，导致训练误差较小但泛化误差较大的问题，即出现欠拟合现象。

为了解决这个问题，引入了 K 折交叉验证方法。该方法将训练样本集划分为 K 个互斥的子集，每次使用其中 K-1 个子集来训练模型，然后用剩余的一个子集来验证模型的性能。重复这个过程 K 次，确保每个子集都被用作验证样本，从而评估决策树的泛化能力。

## 4 事故严重程度划分

安全风险评估的结果涵盖了大量不同运行场景及其相关的安全风险指标信息。为辅助调度人员快速把握安全风险评估的总体情况，可以采用严重程度分级策略对各个运行场景的安全风险进行分类。目前常见的严重程度分级方法包括基于负荷损失量、严重度函数和控制代价等不同原则的方法。

基于负荷损失量的分级方法首先量化系统在发生预期事故后可能损失的负荷量，然后根据不同损失量的大小对各个运行场景的安全风险程度进行排序和分级。其中，文献[14]依据负荷损失量分别划分为 2%~5%、5%~10%以及 10%以上三个级别，来评估各运行场景的安全风险程度。该方法将事故后系统可能的负荷损失量作为安全风险的综合评估指标，同时在计算负荷损失时考虑电网静态和动态安全约束的实际需求。

基于严重度函数的分级方法一般针对特定的安全风险属性。首先，评估安全风险属性的量化指标结果，然后建立相应的严重度函数并计算其数值，最后，设定分级阈值将安全风险程度划分为不同的严重度等级。其中，文献[15]利用基于临界故障清除时间的暂态稳定裕度指标作为评估指标，基于效应理论采用指数函数建立严重度函数，并据此设定 5 个不同等级的分级阈值；文献[16]则以基于电压偏移值和偏移持续时间定义的暂态电压稳定裕度指标为基础，同样采用效应理论中的指数函数建立严重度函数，设定了相应的 5 个等级

分级阈值。这种方法能够反映安全风险增大时严重度增加速度加快的特性，但需要注意，严重度函数的形式和分级阈值往往受人为主观设定影响较大。此外，文献[17]提出了基于潮流计算公式的安全信息指标（SII）来评估系统的静态安全性，并根据 SII 设定了 4 个等级的分级阈值。文献[18]则直接以负荷裕度作为系统静态电压稳定的严重度函数指标，将静态电压稳定的严重度分为 4 个等级。

基于控制代价的分级方法首先确保系统安全所需的控制措施类型，然后根据这些措施的实施代价来划分安全风险的严重度等级，代价越高则对应的等级越高。文献[19]研究了交直流混联电网中典型预防控制措施的控制代价差异，依据预防控制措施的类型将安全风险分为 5 个等级。在此基础上，文献[20]扩展至包括紧急控制措施，根据需要采取的控制措施类型将安全风险分为 6 个等级。文献[19]和[20]同时综合考虑功角、电压和频率等多属性动态安全约束的关键断面最大输电能力，作为安全风险的综合评估指标。通过比较关键断面实际传输功率与其最大输电能力的相对大小来评估系统的安全性。在电网实际运行中，关键输电断面反映了电网运行中的安全薄弱环节，是安全监控的重点。采用多属性动态安全约束的关键断面最大输电能力作为安全风险评估指标，能够全面反映电网运行状态的安全性，符合电网调度运行的实际需求。文献[21]针对风电爬坡事件后的功率平衡问题，依据所需的功率控制措施类型将爬坡事件划分为 5 个严重度等级。文献[22]则针对静态安全风险预警问题，考虑到发电机调整和负荷切除等措施，根据控制措施的需要将安全风险划分为 4 个等级。

在安全风险评估结果的基础上，严重度分级是安全风险预警的最后环节。基于负荷损失量和严重度函数的分级方法的分级阈值往往由电网调度人员根据经验主观设定，虽然能有效反映运行场景的不安全程度，但由于没有对不安全运行场景的可控性进行评估，所以分级结果难以对后续的安全风险防控决策提供更多有价值的决策信息。

## 5 基于人工智能的动态安全评估

电力系统的动态安全评估是保障电力系统安全运行的关键环节之一[23]。电力系统的稳定性指的是在受到扰动或异常情况时，系统能够保持正常运行状态或迅速恢复到稳定状态的能力[24]。稳定性评估通常包括电压稳定性、频率稳定性和功角稳定性[25]。准确的稳定性评估能够帮助运营商及时识别潜在风险，采取预防措施，确保电力系统的安全可靠运行。人工智能技术通过数据驱动的方法和先进的算法，为系统稳定性评估提供了新的思路和工具，显著提升了评估的准确性和效率[26-28]。

### 5.1 基于 AI 的动态安全评估原理

传统的电力系统动态安全评估方法主要依赖于物理模型和经验规则，这些方法在面对复杂和多变的电力系统时存在一定的局限性。人工智能技术通过数据驱动的方法，充分利用海量的监测数据和历史运行数据，自动提取关键特征和模式，从而实现更加准确和高效的稳定性评估[29-32]。

在数据驱动的稳定性评估中，机器学习算法扮演了重要角色。这些算法能够分析电力系统运行数据，识别影响系统稳定性的关键因素，并建立稳定性评估模型。通过学习大量的历史数据，机器学习模型可以预测系统在不同扰动下的稳定性状态，从而帮助识别潜在的风险。人工智能技术不仅可以处理结构化数据，还可以处理非结构化数据，进一步提高了稳定性评估的全面性和准确性。

深度学习模型在处理复杂和高维度数据方面具有显著优势。通过使用深度神经网络，这些模型能够捕捉电力系统运行中的复杂非线性关系，识别出微小但关键的变化，从而提供更细致和精准的稳定性评估[33-36]。特别是在时间序列数据分析中，深度学习模型可以预测系统的未来稳定性变化，为预防性维护和紧急响应提供重要参考[37-39]。

通过数据驱动的方法，电力系统的稳定性评估不再局限于传统的物理模型和专家知识，而是可以动态地适应系统的实时变化[40-45]。随着电力系统运行环境和条件的不断变化，人工智能技术能够不断更新和优化评估模型，确保评估结果的及时性和准确性。这种动态评估能力对于应对电力系统运行中的各种不确定性和突发情况具有重要意义。

### 5.2 AI 在动态安全评估的应用

#### 5.2.1 电压稳定评估

电压稳定性是指电力系统在负荷变化或故障

发生时，能够维持电压在安全范围内的能力[46]。人工智能技术通过对历史运行数据和实时监测数据的分析，可以准确预测电压稳定性，识别可能的电压崩溃点。支持向量机（SVM）是一种有效的监督学习算法，适用于分类和回归问题。在电压稳定性评估中，SVM 可以通过对历史数据的训练，建立电压稳定性预测模型，识别可能导致电压不稳定的运行状态[47-50]。

#### 5.2.2 暂态（功角）稳定评估

暂态稳定性涉及电力系统在大扰动（如短路故障或大负荷变化）后的暂态响应能力[51-52]。深度学习技术可以通过对系统动态响应数据的学习，建立动态稳定性预测模型，评估系统在不同扰动条件下的稳定性。长短期记忆网络（LSTM）[53]和卷积神经网络（CNN）[54]在处理电力系统时序数据和空间数据方面表现突出，能够准确预测系统在不同扰动条件下的动态响应和稳定性。

#### 5.2.3 频率稳定评估

频率稳定性是指电力系统在负荷和发电平衡被打破后，能够迅速恢复到正常频率的能力[55]。人工智能技术通过实时监测和分析系统频率变化，可以提前预测频率不稳定情况，提供相应的调控策略[56-57]。强化学习是一种基于奖励机制的学习算法，适用于动态决策问题[58-60]。在电力系统稳定性评估中，强化学习可以通过模拟系统在不同操作策略下的反应，优化系统调控策略，增强系统的动态稳定性和频率稳定性[61-62]。

### 5.3 优势与挑战

人工智能算法具有强大的数据处理能力，能够快速处理和分析大量数据，从而提高分析的效率和准确性，有效解决工程领域中的分类和回归问题[63-66]。机器学习模型凭借其高预测精度，通过学习历史数据，可以准确预测电力系统在不同运行状态下的安全性，帮助运营商提前采取预防措施。这些模型具备强大的自适应能力，能根据新数据和环境变化进行持续的优化，提高对未知情况的适应能力，从而增强电力系统的安全性和稳定性。机器学习技术还提供了智能故障预测、优化运行调度和安全域评估等多种解决方案，显著提升了电力系统的智能化水平[67-69]。

然而，机器学习模型对数据的质量和数量有较高要求，模型的性能依赖于大量高质量的训练数据。鉴于电力系统数据常伴有噪声和缺失，因此需要有效的数据预处理方法[70-71]。此外，随着信息攻击，如虚假数据注入攻击的日益严重[72-74]，这些基于人工智能的电力系统稳定分析方法面临新的挑战[75-76]。同时，机器学习模型的可解释性亟待提高。深度学习等复杂模型的"黑箱"性质使得其难以解释，这对安全性和可信度构成挑战[77]。因此，发展可解释的机器学习方法以增强模型的透明性成为一项重要任务。考虑到电力系统运行的高度实时性要求，机器学习模型需要在短时间内提供准确的分析和预测结果，这对计算能力和算法效率提出了更高要求[78-79]。电力系统数据来源多样，包括监测数据、历史运行数据和外部环境数据等，如何有效融合这些多源数据以提升模型的整体性能，仍是一个重要挑战[80]。

## 6 结论

基于人工智能的电力系统安全稳定分析能够快速辨识未来安全风险并指导防控决策，是保障新型电力系统安全稳定运行的关键技术。本文从人工智能的视角梳理了电力系统安全稳定分析的所涉及的关键环节，主要包括运行场景生成与缩减、预想事故筛选、安全风险评估和严重程度分级。并针对上述环节综述了目前的研究现状，重点介绍了基于决策树的静态稳定评估方法，并对现有研究中存在的问题及难点进行了分析。在系统动态安全评估方面，人工智能技术提供了更加精确和动态的分析手段。神经网络和深度学习模型通过对大量数据的深入学习，能够准确预测系统在不同运行条件下的稳定性，并据此优化调控策略，从而显著提高系统的整体稳定性和可靠性。

综上所述，人工智能技术在电力系统的安全稳定性分析中的应用，为提升系统的安全性和运行效率提供了强有力的技术保障。随着技术的不断进步和应用的深入，人工智能预计将在电力系统安全管理中发挥更为重要的作用，推动电力系统向更加智能化和高效化的方向发展。